\documentclass[aps,prl,twocolumn,showpacs,preprintnumbers,amsmath,amssymb]{revtex4}

\usepackage{epsfig}
\usepackage{graphics}
\usepackage{dcolumn}
\usepackage{bm}
\usepackage{color} 

\newcommand{\kc}{\color{black}}

\begin{document}


\title{Natural Fueling of a Tokamak Fusion Reactor}
\author{Weigang Wan}
\author{Scott E. Parker}
\author{Yang Chen}
\author{Francis W. Perkins}
\affiliation{Department of Physics, University of Colorado, Boulder, CO 80309, USA}

\date{\today}

\begin{abstract}
A natural fueling mechanism that helps to maintain the main core deuterium and tritium (DT) density
 profiles in a tokamak fusion reactor is discussed. In {\it H}-mode plasmas dominated by ion-temperature
 gradient (ITG) driven turbulence, cold DT ions near the edge will naturally pinch 
radially inward towards the core. This mechanism is due to the quasi-neutral heat flux dominated nature 
of ITG turbulence and still applies when trapped and passing kinetic electron effects are included. 
Fueling using shallow pellet injection or supersonic gas jets is augmented by an inward pinch of could DT fuel. 
The natural fueling mechanism is demonstrated using the three-dimensional toroidal electromagnetic 
gyrokinetic turbulence code {\small GEM} and is analyzed using quasilinear theory. Profiles similar 
to those used for conservative ITER transport modeling that have a completely flat density profile are 
examined and it is found that natural fueling actually reduces the linear growth rates and energy 
transport.
\end{abstract}

\maketitle

The fueling of a tokamak fusion reactor is an important scientific and technological problem~\cite{Parks06,Loarte07,ITER99}. In ITER, deep central fueling 
is somewhat uncertain using present-day high-field side pellet injection schemes~\cite{Pegourie07}. 
Gas puffing and supersonic molecular beam injection~\cite{Yao98} typically fuel the edge~\cite{Loarte07}. In a Tokamak with deuterium, tritium and 
helium ash ion species, it was shown~\cite{Estrada-Mila05} that above certain level of density concentration, the helium ash naturally diffuses outward, helping in ash removal.
Additionally, due to quasi-neutrality, the main DT species go radially inward (or pinch) to balance the helium ash outward flux. 
In this paper, we show how adding cold fuel towards the edge can fuel the core and maintain the DT density profiles. Global multi-species gyrokinetic simulations using the 
{\small GEM} code~\cite{Chen07}, show the effect that cold DT fuel added near the edge will naturally pinch towards the core. The main DT inward fluxes can be significantly reduced by fueling, while the outward flux of helium ash is maintained. The cold DT fuel will heat up due to ion-ion collisions as it moves towards the core, assuming the equilibration time is not too fast, the core density profiles may be 
sustainable with shallow pellet injection.  We will show later that the transport time scale of the cold fuel is faster than the equilibration time in which it heats. 

Two different plasma profiles are studied here: the ``Cyclone DIII-D base case''~\cite{Dimits00,Parker99} and a second case with ITER-relevant density and 
temperature profiles~\cite{Budny08}. A quasi-linear theory is presented to explain the direction of particle fluxes and the fueling mechanism. 
For the ITER-like case, natural fueling has an additional benefit in that it is linearly stabilizing and reduces the energy transport of the main DT.
We first present the simulation results for a Cyclone base case~\cite{Dimits00,Parker99}, which are parameters from a 
DIII-D {\it H}-mode assuming concentric circle flux-surface toroidal geometry. The ``Cyclone base case'' is a standard case 
widely used for benchmarking and for studying physics of turbulent transport in {\it H}-mode plasmas~\cite{Dimits00}. 
Here, we examine a global version of the Cyclone base case~\cite{Parker99}, {\it with} kinetic electrons, and in the 
electrostatic limit.
Recent studies~\cite{Angioni03} show that increased collisionality can reduce the anomalous particle pinch particle transport. We assume a collisionless plasma focusing on parameters towards the core region of a high temperature fusion reactor and the effects of collisionality will be investigated in a future publication.

\begin{figure}
\epsfig{file=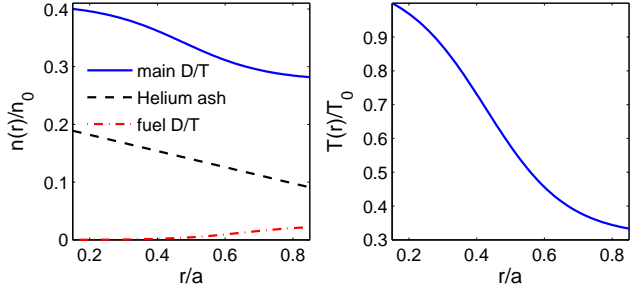,width=8.5cm}
\caption{\label{helpro}(Color online). Density and temperature profiles of the Cyclone base case. Left: densities of main DT, helium ash and the DT fuel; right: temperature of the main species.}
\end{figure}

For the case presented, there are now five ion species: the
 main DT, the helium ash, 
and the DT fuel. The physical parameters are the major radius $R=445\rho_s$ and minor radius $a=160\rho_s$, where 
$\rho_s\equiv\sqrt{T_e/T_i}\rho_i$ and $\rho_i$ 
is the thermal proton gyroradius at a reference temperature $T_0$. The $q$ profile is $q(r)=0.85+2.2\times(r/a)^2$. The 
density $n(r)$ and temperature $T(r)$ profiles, normalized to the reference density $n_0$ and temperature $T_0$, are shown in Fig.~\ref{helpro}. The main DT profiles are selected so that $L_{n,T}^{-1}(r)=B_{n,T}\text{sech}^2[(r-r_0)/w]$, where $L_n\equiv-n/(\partial n/\partial r)$, 
$L_T\equiv-T/(\partial T/\partial r)$, $r_0=a/2$ and $w=40\rho_s$. The constants $B_{n,T}$ are determined by requiring 
$R/|L_n(r_0)|=2.2$ and $R/|L_T(r_0)|=6.9$. 
We assume the helium ash and electrons have the same temperature as the main DT. A linear density profile is assumed for the helium ash, 
as shown in the figure. 
The density profile of the DT fuel has the form $n(r)=C\{1+\tanh[5(r/a-0.65)]\}/4$, in which $C$ denotes the density concentration, e.g., 
$C=0.05$ means the 
peak density of the cold DT fuel is $5\%$ of peak density of main core DT.
 This cold fuel is chosen so that the peak of the fuel density is towards the edge of the simulation 
domain, at $r/a=0.8$.
It is important to note that this is a multi-component plasma and that the 
total DT profile is monotonically decreasing with minor radius throughout 
the simulation domain.

 The temperature of the DT fuel is set to be constant 
and is equal to the temperature of the main species at the simulation boundary near to the edge. 
Hence, the DT fuel is cold compared to the main DT species. 
The three dimensional (3D) simulation domain of $0.15\le r/a \le 0.85$ is discretized using a $128\times 64\times 32$ grid. The time step 
is $\Delta t=2\Omega_i^{-1}$, where $\Omega_i$ is the proton gyrofrequency at $T_0$. We use $4,194,304$ particles per species with realistic mass ratios, e.g., the helium 
mass is $4\times 1837$ times of the electron mass. The simulation uses {\small GEM}, which is a global gyrokinetic $\delta\!f$ particle-in-cell code~\cite{Chen03,Chen07}. 

\begin{figure}
\epsfig{file=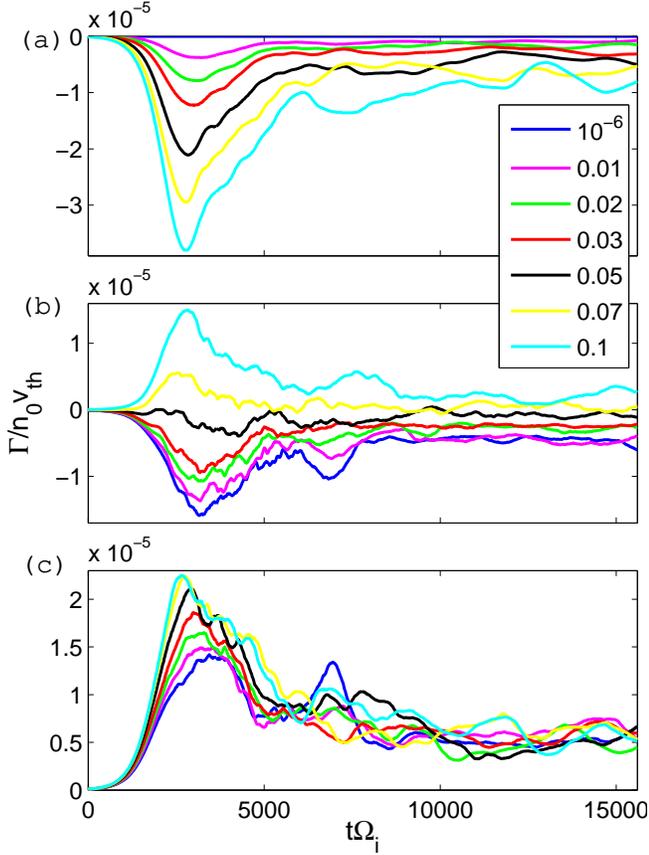,clip=true,width=8.5cm}
\caption{\label{bighel}(Color online). The Cyclone base case: the volume averaged particle fluxes of (a) the deuterium fuel, (b) 
the main deuterium, and (c) the helium ash for $C=10^{-6}$, $0.01$, $0.02$, $0.03$, $0.05$, $0.07$ and $0.1$. The flux is normalized to $n_0 v_\text{th}$, where $v_\text{th}$ is the proton thermal velocity at $T_0$.}
\end{figure}

We study the effects of natural fueling by running simulations with different concentrations of the DT fuel density. 
Figure~\ref{bighel} shows the the particle fluxes of the deuterium fuel, main deuterium, and the helium ash for concentrations 
from $C=10^{-6}$ (no fueling) to $C=0.1$. The particle fluxes are calculated as $\Gamma=\frac{n_0}{N}\sum_j w_j v_{Ex}$, where $w_j$ is the weight of particle $j$, $v_{Ex}$ is the particle's $\bm{E}\times\bm{B}$ drift 
velocity in the $r$ direction, and $N$ is total number of particles. The fluxes are saved at $8$ points along the minor radius $r$, 
and the results shown in Fig.~\ref{bighel} are volume averaged over all the points. The main and fueling tritium fluxes are similar to 
those of the deuterium for this case and not shown. From Fig.~\ref{bighel}(a), the cold fuel flow always goes inward, and the level
increases for bigger $C$. Hence, ``natural fueling,'' that is, the inward pinch of the cold fuel, is demonstrated for this simplified case. 

\begin{figure}[htp]
\epsfig{file=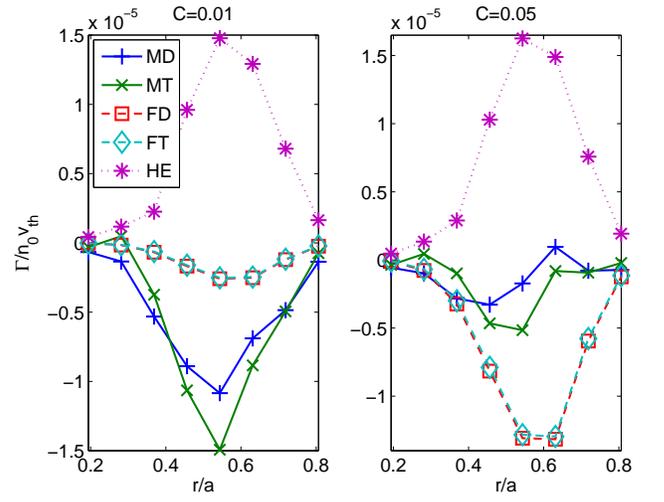,width=8.5cm}
\caption{\label{sflux}(Color online). The time-averaged particle fluxes of the main deuterium (MD), main tritium (MT), fuel deuterium (FD), fuel 
tritium (FT) and helium ash (HE) along $r$ for $C=0.01$ and $C=0.05$.}
\end{figure}

\begin{figure}[htp]
\epsfig{file=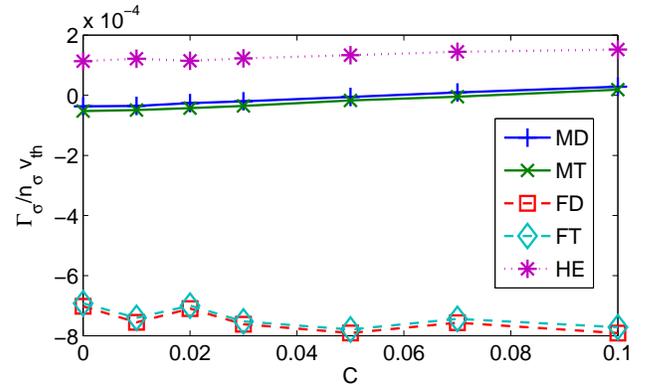,width=8.5cm}
\caption{\label{dhel}(Color online). The density-normalized particle fluxes $\Gamma_\sigma/n_\sigma v_\text{th}$ of the five species for different $C$'s.}
\end{figure}

\citet{Estrada-Mila05} predicts using both theory and gyrokinetic simulation that for $L_{n\text{He}}/L_{ne}<0.84$, the helium ash flow 
should go outward, and this condition is well satisfied for the density profiles presented here. 
For the results in this paper, the helium ash always goes out. Assuming nearly adiabatic electrons, consistent with the properties of ITG
turbulence, quasi-neutrality requires that the out-going helium ash flow must generate some inward going DT flow. Without the DT fuel, 
i.e., for the $C=10^{-6}$ case, the particle fluxes of main DT go inward as the helium ash goes out. This phenomenon is clearly 
shown in Fig.~\ref{bighel}(b) and (c). Therefore, edge fueling is needed to maintain the main DT density profile.

By adding the cold fuel from the edge, the inward particle flux of the main DT is reduced. As shown in Fig.~\ref{bighel}(b), 
for $C=0.05$, the particle flux of the main deuterium is nearly zero. Now, the helium ash goes out at the expense of the cold DT fuel 
instead of the hot main DT. Natural fueling is demonstrated and the main DT profile is maintained. The cold DT fuel, of course,
must be heated as it migrates towards the core. 

This process is better demonstrated in Fig.~\ref{sflux}, where the time-averaged particle fluxes of all five species are shown across 
the minor radius. Comparing the results of $C=0.05$ to $C=0.01$, the increased DT fuel fluxes apparently cancel that of the main DT. 
We note that although this cancellation is significant, it is not complete. This is because the positions of the maximum flux are 
different for the main and fuel DT, due to their individual density profiles. 
The cold fuel concentration density should be kept below a threshold value above which the main DT flux becomes radially outward. As in Fig.~\ref{bighel}(b), for the case $C=0.07$ and $C=0.1$, the main DT fluxes go outward in this case. 

Figure~\ref{dhel} shows the 
effects of fueling on all species. After normalizing by their individual densities, it is clear to see that fueling is most effective at
pinching DT fuel inward. The particle flow of the main DT is altered from inward to near zero. The out-going helium ash flow appears insensitive to 
fueling. This is important, in that the natural fueling does not increase helium ash build up.

A rather crude assumption made in these simulations is that the cold fuel density is given
 and that it's temperature profile is constant.
For this natural fueling mechanism to work, the fuel should remain cold as it migrates towards the core.
The fuel temperature will come in equilibrium with the core temperature on the ion-ion 
equilibration time.
For a Tokamak fusion reactor at $15$ keV in the core such as ITER, assuming the cold fuel temperature is $2$ keV, the ion-ion thermal equilibration time is about $0.03$ second. From Fig.~\ref{dhel}, the averaged flow velocity of the fuel is about $7\times 10^{-4}v_\text{th}$, where $v_\text{th}$ is the proton 
thermal velocity calculated at $T_0$. At fusion temperature, $v_\text{th}$ is more than $10^6$ m/s, 
therefore during the equilibration time, the fuel can go as far as $20$m,
 which is much larger than 
the minor radius. 
 The strong inward flux velocity is not simply caused 
by the negative density gradient of the fuel but rather due to the cold temperature. In fact, a 
pinch exists for cold fuel even in simulations with no density gradient.  Indicating a convective pinch rather
 than diffusive density transport.

Next, a multi-species quasi-linear theory~\cite{Estrada-Mila05} is used to explain the
natural inward pinch of the DT fuel. 
For simplicity, assume there are only two ion species: the hot main deuterium labeled as `$i$', and the cold fuel deuterium, 
as `$I$'. The linear dispersion relation 
is~\cite{Estrada-Mila05} 
\begin{equation}
R_i+R_I=1-i\delta,
\label{dispersion}
\end{equation}
in which the $i\delta$ denotes the non-adiabatic electron response,
\begin{equation}
R_\sigma=f_\sigma\left[\frac{\omega_{*n\sigma}-\omega_d}{\omega}-\frac{\omega_{d}\omega_{*p\sigma}}{\omega^2}+O(k_\theta\rho_s)^2\right],
\end{equation}
where $k_\theta=nq/r$, $n$ is the toroidal mode number, $\sigma=i, I$. $f_I=\epsilon$, $f_i=1-\epsilon$, and $\epsilon\equiv n_I/n_e$. The parameter $\omega_d=2k_\theta\rho_s/R$ is a constant, $\omega_{*n\sigma}=k_\theta\rho_s/L_{n\sigma}$, $\omega_{*T\sigma}=k_\theta\rho_s/L_{T\sigma}$, and $\omega_{*p\sigma}=\omega_{*n\sigma}+\omega_{*T\sigma}$. The quasi-linear particle flux given by
\begin{equation}
\Gamma_\sigma=\text{Re}[ik_\theta\rho_s|\phi|^2R_\sigma(\omega)n_e]
\label{flux}
\end{equation}
is proportional to $|\phi|^2$ at the nonlinear saturation level and the direction of the fuel flux is determined by $-\text{Im}(R_I)$, 
with the inflow of the fuel corresponding to a negative $\Gamma_I$.

The following conditions are generally satisfied in a tokamak plasma, and are assumed in this study: (i) $\omega_{*Ti}>0$, meaning 
the main ion temperature is higher towards the core. (ii) The theory allows that for some Tokamak profiles, like in ITER~\cite{Budny08}, 
$\omega_{*ni}$ could be negative. However, we require $\omega_{*pi}=\omega_{*ni}+\omega_{*Ti}>0$, so the temperature gradient should 
dominate. (iii) The fuel ion density should concentrate at the edge, therefore $\omega_{*nI}<0$. (iv) Since the fuel ions are cold, 
its temperature gradient should be small compared to its density gradient, and we require that $\omega_{*pI}=\omega_{*nI}+\omega_{*TI}<0$. 
Conditions (iii) and (iv) are essential for fueling. We can simply assume $\omega_{*TI}=0$, as in simulations the fuel ion 
temperature is set to be the main ion temperature at the outer boundary point. 

Neglect the $O(k_\theta\rho_s)^2$ term and define $\omega_N\equiv (1-\epsilon)\omega_{*ni}+\epsilon\omega_{*nI}$, 
$\omega_P\equiv (1-\epsilon)\omega_{*pi}+\epsilon\omega_{*pI}$, from 
Eq.~(\ref{dispersion}) we have $\text{Im}(1/\omega^2)=[(\omega_N-\omega_d)\text{Im}(1/\omega)+\delta]/(\omega_d\omega_P)$. With $\text{Im}(1/\omega)=-\gamma/|\omega|^2$ where $\gamma$ is the linear growth rate, the $-\text{Im}(R_I)$ term in Eq.~(\ref{flux}) becomes
\begin{eqnarray}
-\text{Im}(R_I)&=&\frac{\epsilon\omega_{*pI}}{\omega_P}\delta-\omega_d\left(1-\frac{\omega_{*pI}}{\omega_P}\right)\epsilon\frac{\gamma}{|\omega|^2}\nonumber\\
&&+\left(\omega_{*nI}-\frac{\omega_{*pI}}{\omega_P}\omega_N\right)\epsilon\frac{\gamma}{|\omega|^2}.
\label{fuelflux}
\end{eqnarray}

Since $\epsilon$ is small, it is reasonable that $\omega_P>0$. With the negative $\omega_{*pI}$, the first term on the right hand 
side of Eq.~(\ref{fuelflux}), which is proportional to $\delta$, is therefore negative, meaning a non-adiabatic electron out 
flux should enhance fueling. The second term is proportional to $\omega_d$ which comes from the toroidal geometry of the device, and 
this term is also negative, so toroidal effects are also favorable for natural fueling. In the limit of $\omega_{*TI}=0$, the third term 
becomes $\epsilon(1-\epsilon)\omega_{*nI}\omega_{*Ti}\gamma/(|\omega|^2\omega_P)$. Because of the density profile
and the temperature gradient of the main ions, this term is also negative. We conclude that the fuel flux must be naturally negative and 
going inward towards the core. 

Due to quasi-neutrality, adding a negative fueling flux should increase the main ion particle transport. This effect can also be seen 
by switching $I\rightarrow i$ and $\epsilon\rightarrow 1-\epsilon$ in Eq.~(\ref{fuelflux}), in which all three terms on the right-hand-side 
are positive for the main ion, satisfying $\text{Im}(R_i)+\text{Im}(R_I)=-\delta$. In the case where only two ion species are present, e.g., 
hot main deuterium and cold fuel deuterium, the fuel ion goes in at the expense of hot ions going out, and therefore the natural fueling 
is of no value. However, when helium ash is present, the main DT pinch as the helium ash goes 
outward. Fueling is then useful to maintain the main DT profile, and balance the outgoing helium ash flux.

\begin{figure}[htp]
\epsfig{file=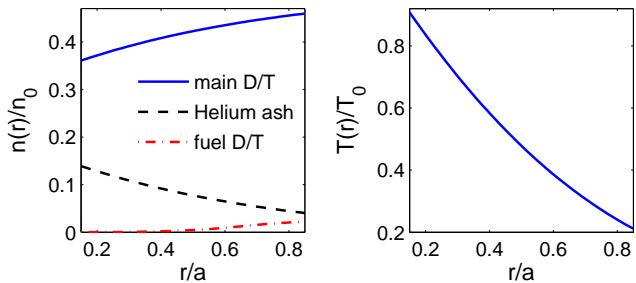,width=8.5cm}
\caption{\label{iterpro}(Color online). Profiles of the ITER-like case.}
\end{figure}
\begin{figure}[htp]
\epsfig{file=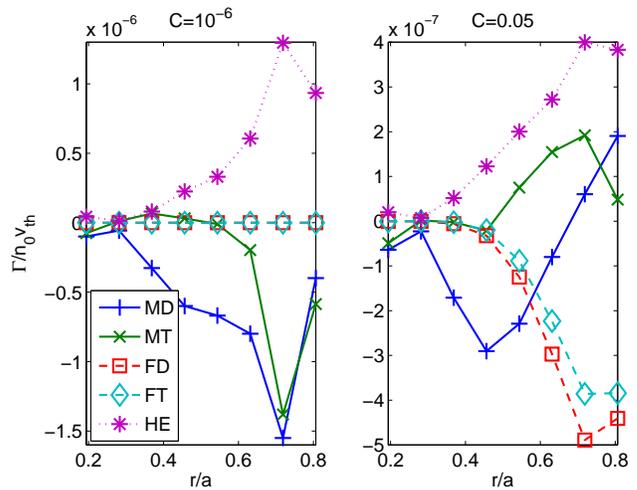,width=8.5cm}
\caption{\label{sfiter}(Color online). The ITER-like case: the particle fluxes for $C=10^{-6}$ and $0.05$.}
\end{figure}

The global version of the Cyclone base case presented here have monotonically decreasing density profiles that peak at the magnetic axis.
 So far there's no general consensus on the ITER density profile. 
One rather conservative study~\cite{Budny08} shows ITER {\it H}-mode profiles with a completely flat density.
The inward peaking helium ash then causes the main DT density to be hollow, or peak near the edge. 
We now examine
the natural fueling mechanism in this situation. 
The next set of simulations have the same physical parameters as in the Cyclone base case, but use the ITER-like hollow DT density and temperature 
profiles, as shown in Fig.~\ref{iterpro}.
 
The simulation results for this flat density ITER-like case are shown in Fig.~\ref{sfiter}.
This completely flat density profile is overall much less unstable and 
as a result the cold fuel flux is also lower. This reduces the travel length of the fuel to about $0.8$m 
before they are heated by ion-ion thermal equilibration. \kc The outward going helium ash flow still 
causes the main DT to go inward without fueling, but there is a different phenomenon. 
The main deuterium flux 
now separates from the main tritium flux. Similar to the previous case, adding cold DT fuel reduces 
the inward particle pinch of the main DT. 
 However, due to the separation of the main D and T, it is very easy to overshoot with too large a
 $C$, thereby, causing
the main tritium to go outward. This overshoot is shown 
in the right panel of Fig.~\ref{sflux}. Given this effect of DT separation in the ITER-like case, 
one may consider using
different density profiles for the deuterium fuel and tritium, possibly less tritium and more 
deuterium, but it is beyond the scope of 
this Letter and will require future investigations.

\begin{figure}[htp]
\epsfig{file=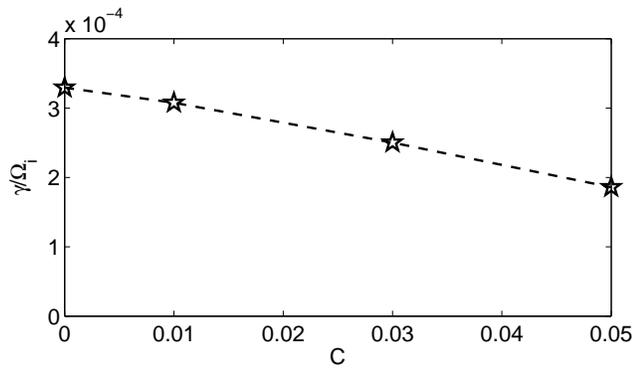,width=8.5cm}
\caption{\label{cgp}Linear growth rates for the ITER-like case.}
\end{figure}

Another significant new result of the ITER-like case presented here is that the linear growth rate of the ITG-induced instability is 
reduced by natural fueling, as shown in Fig.~\ref{cgp}. This bonus is due to the hollow density profile of the main DT. 
In general, at low real frequency the linear growth rate $\gamma$ from Eq.~(\ref{dispersion}) is roughly $\sqrt{\omega_d\omega_P}$ 
with fueling and $\sqrt{\omega_d\omega_{*pi}}$ for with main ions only. Since $0<\omega_P<\omega_{*pi}$, fueling decreases $\gamma$ and 
hence is a stabilizing effect. This effect is insignificant for the Cyclone base case: from $C=10^{-6}$ to $C=0.1$, $\gamma$ 
is only reduced by $2\%$, at about $6.8\times 10^{-4}\Omega_i$. However, in the ITER-like case, since $\omega_{*ni}$ is negative, for 
comparable temperature profiles, $1/\omega_{*pi}$ is bigger than the Cyclone base case and hence 
$\omega_P/\omega_{*pi}=1-\epsilon+\epsilon\omega_{*pI}/\omega_{*pi}$ is smaller. Therefore $\gamma$ is reduced more by fueling. 
As a result, the heat fluxes of both main DT and helium ash are also reduced. For $C=0.05$ in the ITER-like case, the heat fluxes 
are reduced by nearly $50\%$ from $C=10^{-6}$. 

We thank Glenn Bateman, Lehigh Univ. for insightful discussions.
This work is supported by the Department of Energy Scientific Discovery through Advanced Computing Program, 
Center for Plasma Edge Simulation 
and Center for the Study of Plasma Microturbulence. 


\begin{thebibliography}{12}
\expandafter\ifx\csname natexlab\endcsname\relax\def\natexlab#1{#1}\fi
\expandafter\ifx\csname bibnamefont\endcsname\relax
  \def\bibnamefont#1{#1}\fi
\expandafter\ifx\csname bibfnamefont\endcsname\relax
  \def\bibfnamefont#1{#1}\fi
\expandafter\ifx\csname citenamefont\endcsname\relax
  \def\citenamefont#1{#1}\fi
\expandafter\ifx\csname url\endcsname\relax
  \def\url#1{\texttt{#1}}\fi
\expandafter\ifx\csname urlprefix\endcsname\relax\def\urlprefix{URL }\fi
\providecommand{\bibinfo}[2]{#2}
\providecommand{\eprint}[2][]{\url{#2}}

\bibitem[{\citenamefont{Parks and Perkins}(2006)}]{Parks06}
\bibinfo{author}{\bibfnamefont{P.~B.} \bibnamefont{Parks}} \bibnamefont{and}
  \bibinfo{author}{\bibfnamefont{F.~W.} \bibnamefont{Perkins}},
  \bibinfo{journal}{Nucl. Fusion} \textbf{\bibinfo{volume}{46}},
  \bibinfo{pages}{770} (\bibinfo{year}{2006}).

\bibitem[{\citenamefont{Loarte et~al.}(2007)}]{Loarte07}
\bibinfo{author}{\bibfnamefont{A.}~\bibnamefont{Loarte}} \bibnamefont{et~al.},
  \bibinfo{journal}{Nucl. Fusion} \textbf{\bibinfo{volume}{47}},
  \bibinfo{pages}{S203} (\bibinfo{year}{2007}).

\bibitem[{\citenamefont{{ITER Physics Expert Group on Divertor}
  et~al.}(1999)}]{ITER99}
\bibinfo{author}{\bibnamefont{{ITER Physics Expert Group on Divertor}}}
  \bibnamefont{et~al.}, \bibinfo{journal}{Nucl. Fusion}
  \textbf{\bibinfo{volume}{39}}, \bibinfo{pages}{2391} (\bibinfo{year}{1999}).

\bibitem[{\citenamefont{P\'{e}gouri\'{e}}(2007)}]{Pegourie07}
\bibinfo{author}{\bibfnamefont{B.}~\bibnamefont{P\'{e}gouri\'{e}}},
  \bibinfo{journal}{Plasma Phys. Control. Fusion}
  \textbf{\bibinfo{volume}{49}}, \bibinfo{pages}{R87} (\bibinfo{year}{2007}).

\bibitem[{\citenamefont{Yao et~al.}(1998)}]{Yao98}
\bibinfo{author}{\bibfnamefont{L.}~\bibnamefont{Yao}} \bibnamefont{et~al.},
  \bibinfo{journal}{Nucl. Fusion} \textbf{\bibinfo{volume}{38}},
  \bibinfo{pages}{631} (\bibinfo{year}{1998}).

\bibitem[{\citenamefont{{Estrada-Mila}
  et~al.}(2005)\citenamefont{{Estrada-Mila}, Candy, and
  Waltz}}]{Estrada-Mila05}
\bibinfo{author}{\bibfnamefont{C.}~\bibnamefont{{Estrada-Mila}}},
  \bibinfo{author}{\bibfnamefont{J.}~\bibnamefont{Candy}}, \bibnamefont{and}
  \bibinfo{author}{\bibfnamefont{R.~E.} \bibnamefont{Waltz}},
  \bibinfo{journal}{Phys. Plasmas} \textbf{\bibinfo{volume}{12}},
  \bibinfo{pages}{022305} (\bibinfo{year}{2005}).

\bibitem[{\citenamefont{Chen and Parker}(2007)}]{Chen07}
\bibinfo{author}{\bibfnamefont{Y.}~\bibnamefont{Chen}} \bibnamefont{and}
  \bibinfo{author}{\bibfnamefont{S.~E.} \bibnamefont{Parker}},
  \bibinfo{journal}{J. Comput. Phys.} \textbf{\bibinfo{volume}{220}},
  \bibinfo{pages}{839} (\bibinfo{year}{2007}).

\bibitem[{\citenamefont{Parker et~al.}(1999)\citenamefont{Parker, Kim, and
  Chen}}]{Parker99}
\bibinfo{author}{\bibfnamefont{S.~E.} \bibnamefont{Parker}},
  \bibinfo{author}{\bibfnamefont{C.}~\bibnamefont{Kim}}, \bibnamefont{and}
  \bibinfo{author}{\bibfnamefont{Y.}~\bibnamefont{Chen}},
  \bibinfo{journal}{Phys. Plasmas} \textbf{\bibinfo{volume}{6}},
  \bibinfo{pages}{1709} (\bibinfo{year}{1999}).

\bibitem[{\citenamefont{Dimits et~al.}(2000)}]{Dimits00}
\bibinfo{author}{\bibfnamefont{A.~M.} \bibnamefont{Dimits}}
  \bibnamefont{et~al.}, \bibinfo{journal}{Phys. Plasmas}
  \textbf{\bibinfo{volume}{7}}, \bibinfo{pages}{969} (\bibinfo{year}{2000}).

\bibitem[{\citenamefont{Budny et~al.}(2008)}]{Budny08}
\bibinfo{author}{\bibfnamefont{R.~V.} \bibnamefont{Budny}}
  \bibnamefont{et~al.}, \bibinfo{journal}{Nucl. Fusion}
  \textbf{\bibinfo{volume}{48}}, \bibinfo{pages}{075005}
  (\bibinfo{year}{2008}).

\bibitem[{\citenamefont{Angioni et~al.}(2003)}]{Angioni03}
\bibinfo{author}{\bibfnamefont{C.}~\bibnamefont{Angioni}} \bibnamefont{et~al.},
  \bibinfo{journal}{Phys. Rev. Lett.} \textbf{\bibinfo{volume}{90}},
  \bibinfo{pages}{205003} (\bibinfo{year}{2003}).

\bibitem[{\citenamefont{Chen and Parker}(2003)}]{Chen03}
\bibinfo{author}{\bibfnamefont{Y.}~\bibnamefont{Chen}} \bibnamefont{and}
  \bibinfo{author}{\bibfnamefont{S.~E.} \bibnamefont{Parker}},
  \bibinfo{journal}{J. Comput. Phys.} \textbf{\bibinfo{volume}{189}},
  \bibinfo{pages}{463} (\bibinfo{year}{2003}).

\end{thebibliography}

\end{document}